\documentclass[3p,shortbib,twocolumn]{elsarticle}

\usepackage{amsmath,amssymb}
\usepackage{graphicx,epstopdf}
\usepackage{graphicx}
\usepackage[usenames,dvipsnames]{xcolor}
\usepackage{slashed}
\usepackage{ulem}
\usepackage[dvipsnames]{xcolor}

\usepackage{hyperref}
\hypersetup{colorlinks=true}

\newcommand{\be}{\begin{equation}}
\newcommand{\ee}{\end{equation}}
\newcommand{\bea}{\begin{eqnarray}}
\newcommand{\eea}{\end{eqnarray}}

\newcommand{\tsf}[1]{\textsf{#1}}

\definecolor{aaaa}{rgb}{0.99, 0.4, 0.01}
\definecolor{bbbb}{rgb}{0.5, 0.3, 0.9}

\newcommand{\Eq}[1]{\mbox{Eq.~(#1)}}

\newcommand{\ud}{\mathrm{d}}

\newcommand{\LCperp}{{\scriptscriptstyle \perp}}

\newcommand{\bi}{\begin{itemize}}
\newcommand{\ei}{\end{itemize}}

\makeatletter
\def\ps@pprintTitle{%
 \let\@oddhead\@empty
 \let\@evenhead\@empty
 \def\@oddfoot{}%
 \let\@evenfoot\@oddfoot}
\makeatother

\begin{document}

\begin{frontmatter}

\title{Toward the observation of interference effects in nonlinear Compton scattering}
\author{A.~Ilderton}
\author{B.~King}
\author{S.~Tang\corref{Cof}}
\cortext[Cof]{Corresponding author}
\ead{suo.tang@plymouth.ac.uk}
\address{Centre for Mathematical Sciences, University of Plymouth, Plymouth, PL4 8AA, United Kingdom}
\begin{abstract}
The photon spectrum from electrons scattering on multiple laser pulses exhibits interference effects not present for scattering on a single pulse. We investigate the conditions required for the experimental observation of these interference effects in electron-laser collisions, in particular analysing the roles of the detector resolution and the transverse divergence of the pump electron beam.
\end{abstract}

\end{frontmatter}

\section{Introduction}
The double-slit experiment famously demonstrates quantum interference effects. An analogous double slit can be made by using intense electromagnetic fields to polarise patches of the quantum vacuum; the resulting all-optical diffraction grating can be realised in space~\cite{King:2013am,King:2013zz}, time~\cite{lindner05,Hebenstreit:2009km,Akkermans:2011yn,Li:2014xga}, and both space and time~\cite{ilderton2019coherent}.  If particles are scattered off the polarised vacuum, they can exhibit double-slit-like interference effects.

The analogy with `material' double-slit patterns holds in the semiclassical approximation~\cite{Akkermans:2011yn,Dumlu:2011rr,Schneider:2018huk}, but exactly solvable models show that interference structure can be much richer~\cite{ilderton2019coherent}. In particular, coherent enhancement of quantum effects can persist despite the semiclassical approximation predicting no interference. This prompts the question of to what extent interference effects can be observed in modern laser-particle collisions. In this paper we will take some steps toward answering this question, considering interference effects in the photon emission spectrum of electrons colliding with laser pulses, known as `nonlinear Compton scattering'~\cite{nikishov64,ritus1985quantum,bula96,seipt2017volkov}.

One motivation for studying nonlinear Compton is that experimental campaigns such as \mbox{LUXE} at \mbox{DESY}~\cite{Hartin:2018sha,abramowicz2019letter} and \mbox{E320} at \mbox{FACET-II}~\cite{slacref1} are planning to investigate the process using conventionally-accelerated electron beams collided with high power laser pulses, thereby reaching a higher level of precision than current state-of-the-art experiments employing laser-wakefield accelerated electrons~\cite{cole18,poder18}. This precision is vital to assess the accuracy of theory and simulation modelling in the high-intensity regime;  perturbative approaches do not suffice at high intensity, and despite recent successes there remain discrepancies between theory and experimental measurements~\cite{cole18,poder18}.

The important aspect here is precision; laser-matter experiments are currently simulated using particle-in-cell codes based on the {locally constant field approximation} (LCFA, for reviews see~\cite{ritus1985quantum,PRE2015Gonoskov}). However, the LCFA is blind to interference effects~\cite{harvey15} because it is, by construction, local, while interference effects originate from accumulated phases as particles pass through a laser pulse. Essentially, the LCFA sums \textit{incoherently} over contributions from different parts of the trajectory~\cite{dinu16}, completely missing \textit{coherent} quantum effects. Precision experiments such as \mbox{LUXE} and \mbox{E320} will however probe regions where the applicability of the LCFA is questionable. An improved understanding of when current numerical schemes fail will therefore be indispensable in analysing future experimental results~\cite{meuren17,PRA2019Ilderton}. Furthermore, as experimental precision improves, so must theoretical predictions. In contrast to previous investigations, we will show how considerations such as detector resolution are central to the measurability of quantum interference phenomena~\cite{PRSTAB040704,Seipt:2013hda,Krajewska:2014fwa,Krajewska_2014,harvey15,Ivanov:2016jzt,JANSEN201771,Titov:2018bgy,Granz:2019sxb,ilderton2019coherent}. As we will see, interference is a precision effect which presents a challenge to theory and experiment.

This paper is organised as follows. In Sec.~\ref{SECT:THEORY} we show how interference effects arise in the spectra of photons emitted from electrons scattering on a sequence of laser pulses. In Sec.~\ref{SECT:OBS} we analyse the conditions necessary for the observation of interference effects. In Sec.~\ref{SECT:EXPT} we discuss potential experimental signals, accounting for beam polarisation and electron bunch size effects. We conclude in Sec.~\ref{SECT:CONCS}.

\section{NLC in plane wave background}~\label{SECT:THEORY}

We consider a simple scenario, in which a high-energy electron (mass $m$, absolute charge $e$) with momentum $p_\mu$ collides with a laser pulse, scatters, and radiates a photon with momentum~$l_\mu$. The extension to a bunch of electrons will be given below. The laser is modelled as a plane wave with wavevector $k_\mu = \omega_0(1,0,0,1)$, central frequency $\omega_0$. The strength and shape of the wave are encoded in the scaled potential $a^{\mu} = eA^{\mu}(\phi)$, in which $\phi=k \cdot x$. The interaction energy is characterised by $\eta = k\cdot p/m^2$.

The calculation of the photon emission probability and spectrum, starting from an S-matrix element with Volkov wavefunctions~\cite{volkov35}, is well documented in the literature, see for {example~\cite{seipt2017volkov} for an introduction}. We therefore present just the final expression for the differential emission probability, parameterised by the three components of the emitted photon momentum: these are $s := k \cdot l/k\cdot p$, the light-front momentum fraction of the photon, with $0<s<1$, and {$\ell^\LCperp := (l^x,l^y)/sm$} the two, normalised, momentum components in the plane perpendicular to the laser propagation direction. $\ell^\LCperp$ is related to the polar ($\theta$) and azimuthal ($\psi$) scattering angles of the photon by
\be
	\ell^\LCperp = \frac{m\eta}{\omega_0}\frac{\sin \theta}{1+\cos \theta}
	\left(\cos \psi,\sin \psi \right) \,.
\label{Eq_scattering_angle}
\ee
In terms of $s$ and $\ell^\LCperp$ the differential emission probability, summed over emitted photon {polarisations~\cite{boca09,seipt2017volkov}, summed (averaged) over spins of the scattered (initial) electron~\cite{king2020nonlinear}, is}
\begin{align}
	\frac{\ud^3{\tsf{P}}}{\ud s~\ud^2\ell^{\LCperp}} &= \frac{\alpha s \big[ g
(SI^{*}+S^{*}I-2 F\cdot F^*)-|I|^2 \big]}{(2\pi\eta)^2 (1-s)}
\label{Eq_NCS_Simplified}
\end{align}
where $g \equiv 1/2+ s^2/[4(1-s)]$ and the functions $I$, $F^\mu$ and $S$ are defined as follows. Let the classical kinetic momentum of the electron in the plane wave background be written
\be
\label{def:pi}
	\pi_{p}(\phi)= p - a(\phi) + \frac{2p\cdot a(\phi) - a^{2}(\phi)}{2k \cdot p}k\,.
\ee
We then define
\be
\label{def:om}
	\Omega_{\gamma}(\phi) = \frac{l\cdot \pi_p(\phi)}{m^2 \eta (1-s)}.
\ee
In terms of \Eq{\ref{def:pi}} and \Eq{\ref{def:om}} we have
\begin{subequations}
\label{Eq_phase_integral}
\begin{align}
	I~&=\int\!\ud\phi \, \bigg[1-\frac{l\cdot \pi_{p}(\phi)}{l\cdot p}\bigg] e^{i\Phi(\phi)}\,, \\
	F^\mu &= \frac{1}{m}\int\!\ud \phi ~ a^\mu(\phi) e^{i\Phi(\phi)}\,, \\
	S~&= \frac{1}{m^2}\int\!\ud\phi~ a(\phi)\cdot a(\phi) e^{i\Phi(\phi)}\,,
\end{align}
\end{subequations}
in which the phase $\Phi(\phi)=\int^{\phi}_{\phi_{i}}d\phi'\Omega_{\gamma}(\phi')$, and $\phi_i$ is the initial phase at which the pulse turns on. Each of the integrals in \Eq{\ref{Eq_phase_integral}}, which occur in the S-matrix element, extends only over the pulse duration~\cite{boca09,ilderton2019absorption}.

\subsection{Two-pulse interference}
%
We now choose the potential $a^\mu$ to describe a sequence of two pulses with zero temporal overlap, 
\be
	a^\mu(\phi) = a^\mu_1(\phi)+a^\mu_2(\phi) \;,
\ee
in which $a_1$ ($a_2$) is nonzero only in the range $\phi_{1i}$ to $\phi_{1f}$ ($\phi_{2i}$ to $\phi_{2f}$), and there is a phase gap $\Delta = \phi_{2i} - \phi_{1f}$ between the two pulses. Inserting this into \Eq{\ref{Eq_phase_integral}}, each of the integrals breaks up into two contributions, one from each of the pulses, but where, crucially, the second contribution comes with an accumulated phase $\Phi_f$, i.e.
\be
	I = I_{1}+e^{i\Phi_f} I_{2}\,,
\label{Eq_diff_pulses}
\ee
%
with $I_{1,2}$ given by \Eq{\ref{Eq_phase_integral}} in terms of $a_{1,2}$. Exactly analogous expressions hold for $F^\mu$ and $S$. The interference phase itself is
\begin{align}
	\Phi_f = \int\limits^{\phi_{2i}}_{\phi_{1i}} \!\ud \phi \, \Omega_{\gamma}(\phi) = \Phi_{1}(\phi_{1f})+ \frac{\Delta\, l\cdot p}{m^2 (1-s)\eta} \;.
\label{Eq_interference_phase}
\end{align}
The first term $\Phi_{1}(\phi_{1f})$ depends on $a_1$ and comes from the interaction with the first pulse. The second term comes from integrating over the phase gap between the pulses, and so depends only on the separation $\Delta$, not on the form of the pulses themselves\footnote{The interference phase is nonzero even if the pulse separation is zero ($\Delta=0$). Mathematically, this reflects the fact that we could describe e.g.~a full cycle of the field as two half-cycles of zero separation. Physically, it reflects the fact that there is interference from different parts of a single pulse with itself~\cite{Heinzl:2010vg,Krajewska:2014fwa,harvey15} -- such effects are here part of the single pulse spectrum, while our interest is in the effect on this spectrum due to  interference with a time-delayed second pulse.}.  Inserting \Eq{\ref{Eq_diff_pulses}} into the differential probability \Eq{\ref{Eq_NCS_Simplified}}, the interference phase appears with cross terms between the two pulses, e.g.~
\be
	|I|^2 = |I_{1}|^{2}+|I_{2}|^2 +e^{i\Phi_f}I^{*}_{1}I_{2}~+e^{-i\Phi_f} I_{1}I^{*}_{2} \,,
\label{Eq_NLC_Two_Interference}
\ee
and similarly for $F^\mu$ and $S$. If we drop these terms, i.e.~neglect interference effects, then the total emission probability reduces to an incoherent sum over contributions from each pulse. If, on the other hand, we consider two identical pulses such that the contributions from each pulse are equivalent, $I_1=I_2$ and so on, then  the differential probability becomes
\begin{align}
	\frac{\ud^3{P}}{\ud s~\ud^2\ell^{\LCperp}}=  2\left(1+\cos\Phi_f\right)  \frac{\ud^3{P}}{\ud s~\ud^2\ell^{\LCperp}}\bigg|_\text{one pulse} \;,
\label{Eq_NLC_Interference_same_prob}
\end{align}
with the one-pulse expression exactly as in \Eq{\ref{Eq_NCS_Simplified}}.

We stress that interference effects are not captured by the LCFA widely employed in particle-in-cell codes to approximate QED processes and model laser-matter experiments. The LCFA for the probability depends only on the local value of the field, and thus is blind to the accumulated phase which carries the interference effects. For the case of two identical pulses as above, the LCFA would simply return twice the LCFA probability for scattering in a single pulse.

\section{Interference and resolution}\label{SECT:OBS}
Observe that the interference phase $\Phi_f$ is real and depends linearly on the pulse separation $\Delta$ (at fixed momentum variables). This implies that interference effects do not decay with an increase of the pulse separation~$\Delta$; we know, though, that decoherence will wash out quantum effects ~\cite{Zurek:2003zz,Schlosshauer:2003zy}.
In writing down the probabilities above (and by extension the related emission spectra) we are assuming not only no further interaction with the environment, but also propagation to infinity and perfect resolution in measuring the spectrum.
Realistically, though, all measurements are limited in their resolution, which we can think of as causing `binning' of the data collected, be it e.g.~photon energy or an angular distribution, and so on.
Furthermore, as we sum over/integrate out momenta, interference effects are typically washed out due to the summation over the oscillations of the interference phase.
For instance, suppose we want to measure the angular photon distribution, as a function of $\ell^\LCperp$, at $s=s_c$. This is simply read off from \Eq{\ref{Eq_NCS_Simplified}} or \Eq{\ref{Eq_NLC_Interference_same_prob}}. However, to allow for a finite detector resolution, we should sum over contributions in some range $\delta s$ centred at $s_c$. For $\delta s  \ll s_c$ and $\delta s  \ll 1-s_c$ one can check from the form of $\Phi_f$ that
\begin{align}
	\int^{s_c+ \delta s/2}_{s_c- \delta s/2} \!\ud s\, e^{i\Phi_f}  \propto \frac{1}{\Delta}\sin\bigg[\frac{\ell_\LCperp^2+1}{4\eta(1-s_c)^2}\Delta \delta s\bigg]\,,
\end{align}
for $\Delta\rightarrow \infty$. Hence  the inclusion of detector resolution results in the expected decay of the interference effect with pulse separation.  

To illustrate this discussion we investigate here the extent to which interference effects persist in the angular photon spectrum at some $s_c$, but with a resolution $\delta s$, i.e.~we study
\begin{align}\label{WTACTUALF}
\int^{s_c+ \delta s/2}_{s_c- \delta s/2} \ud s~\frac{\ud^3\textsf{P}}{\ud s~\ud^2\ell^{\LCperp}}\,.
\end{align}
We consider the case that our separated pulses have precisely the same functional form, so
\begin{align}
	a^\mu_1(\phi)&=m\xi \varepsilon^\mu_{1}\,  \sin\phi~\cos^2\left(\frac{\phi}{4\sigma}\right),
\label{Eq_field}
\end{align}
for $|\phi|<2\pi\sigma$ and zero otherwise, where \mbox{$\varepsilon^\mu_{1}=(0,1,0,0)$} is the linear polarisation vector, $\xi$ is the normalised field amplitude, and \mbox{$a_2(\phi) =a_1(\phi-4\pi \sigma -\Delta)$}, which gives a phase gap of $\Delta$ between the two pulses.

The question then arises as to what parameters are needed to see interference effects. A natural condition on the phase $\Phi_f$ for the persistence of interference effects is that the \textit{change} in $\Phi_f$, over some considered parameter interval (in e.g. $\ell^{\perp}$ or $s$), should be less than $2\pi$. Although the interference effect is periodic with the change in $\Phi_f$, the \textit{relative} size of the effect compared to the incoherent (non-interference) terms \textit{decreases} as the considered interval increases. The explicit expression for $\Phi_f$ in terms of the variables $\ell^\LCperp$ and $s$ is
\begin{align}
	\frac{s(\ell^{2}_{\LCperp}+1)\Delta}{2\eta(1-s)} +\frac{s}{\eta} \int^{\phi_{\textrm{1f}}}_{\phi_{\textrm{1i}}}
\!\ud\phi\, \frac{1+ (\ell^\LCperp+{a}^\LCperp_{1}/m)^2}{2(1-s)},
\end{align}
for a head-on collision, from which we infer that for the observation of interference the two pulses should not be too far apart ($\Delta$ small) and the resolution $\delta s$ should be small. Furthermore, the intensity $\sim a$ should not be too large and the pulse duration $\sim \phi_{\textrm{1f}}-\phi_{\textrm{1i}}$ should be short. The last two conditions confirm that interference effects will be dominant in precisely the regions where the LCFA fails. We proceed to consider short and relatively weak pulses, and small~$s$.

\begin{figure}[t!!]
 \includegraphics[width=0.49\textwidth]{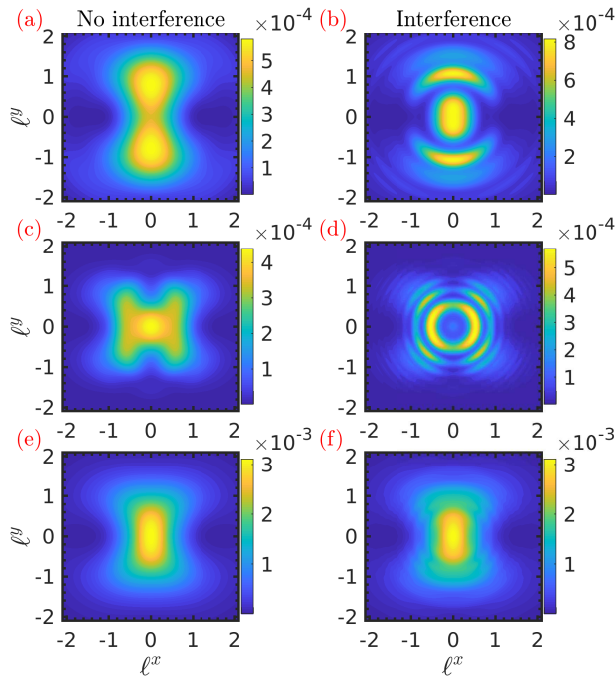}
\caption{Transverse photon distribution \Eq{\ref{WTACTUALF}} from the head-on collision between an $8~\textrm{GeV}$ electron and a one-cycle optical laser, $\omega_0=1.55$~eV, at $\xi=2$, with (right column) and without (left column) interference effects. Upper panels  $s_c=0.07$, $\delta s=0.02$; middle panels $s_c=0.15$, $\delta s=0.02$; lower panels $s_c=0.11$, $\delta s=0.1$. The energy parameter is $\eta=0.095$ and the pulse separation is $\Delta=2\pi$.}
\label{Fig_same_pulse}
\end{figure}

We therefore consider a one-cycle pulse\footnote{We choose such a short pulse simply for demonstration purposes. The effect will persist with longer pulses, but a better resolution will be required for their detection. We remark that calculations based on the approximation of \textit{infinitesimally} short, delta-function pulses, can successfully reproduce experimental interference effects in which femtosecond pulses were employed~\cite{Bigourd:2008zz}.}  with $\xi=2$. The pulse separation is chosen to be $\Delta=2\pi$, also corresponding to one cycle.  In Fig.~\ref{Fig_same_pulse}, we present the transverse distributions of the emitted photons \Eq{\ref{WTACTUALF}}, from the head-on collision of an $8~\textrm{GeV}$ electron (a typical energy scale for \mbox{LUXE} and \mbox{E320} experiments~\cite{abramowicz2019letter,slacref1}), with the two laser pulses\footnote{A small angular deviation $\theta_i$ from a head-on collision between the electron and the laser will not lead to substantial differences. This is because the key parameter, $\eta \propto 1+ \cos\theta_i$, changes only slightly as $\theta_i$ is increased from 0.}. {The spectra are obtained by numerically evaluating \Eq{\ref{Eq_phase_integral}} at many points to integrate over the pulse and inserting the results into \Eq{\ref{Eq_NCS_Simplified}}.} We consider various energy resolutions and also compare (in the left hand column) with the spectra which would be obtained without interference, which is just the incoherent sum of two single-pulse results [i.e.~set $\cos\Phi_f=0$ in \Eq{\ref{Eq_NLC_Interference_same_prob}}].
The upper panels (a) and (b) clearly show the modulation of the spectrum introduced by interference when one has a resolution $\delta s= 0.02$ at small $s_c=0.07$. {As we have high-energy electrons which mainly emit forwards, this range may be mapped directly to an} energy resolution of \mbox{$160$ MeV} around a central energy of \mbox{$560$ MeV}. {(Measurement of such photon energies may be performed e.g. by using a pixelated scintillator \cite{behm2018}.)}
We see, in general, that the position of maxima and minima can change position when interference is taken into account: instead of two symmetric off-centre maxima when interference is neglected in (a) we see three maxima when interference is included at the centre of image (b)~\cite{ilderton2019coherent};  in the middle panels with $\delta s=0.02$ at larger $s_c=0.15$, interference splits the single central broad maximum (c) into a sharp ring structure with a central minimum in (d).

\begin{figure}[t!!]
 \center{\includegraphics[width=0.8\columnwidth]{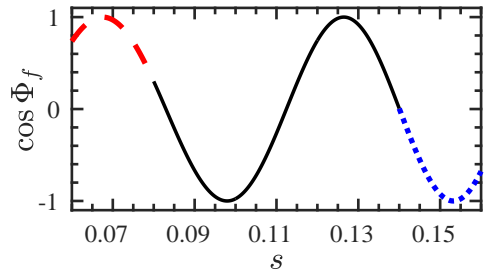}}
\caption{\label{Fig_same_interf_factor} Oscillation of the interference factor $\cos\Phi_f$ at the point $\ell^\LCperp=0$, as $s$ varies. The red/dashed part ($0.06<s<0.08$) and blue/dotted part ($0.14<s<0.16$) lead to (respectively) constructive and deconstructive interference as discussed in the text. All parameters are same as in Fig.~\ref{Fig_same_pulse}.}
\end{figure}

However, with a lower $\delta s = 0.1$ around $s_c=0.11$, corresponding to an energy resolution of $800$~MeV around a central energy of $880$ MeV, the interference effect becomes much weaker, as expected. This is shown in the lower panels; the interference fringes in (f) are weak and the result is almost indistinguishable from the incoherently doubled single-pulse result in~(e).

To explain these features, we plot in Fig.~\ref{Fig_same_interf_factor} the interference factor $\cos\Phi_f$ at the point $\ell^\LCperp=0$, i.e.~at the centre of each of the plots in Fig.~\ref{Fig_same_pulse}.
In the region \mbox{$0.06<s<0.08$} the factor $\cos\Phi_f$ remains close to $1$ ({dashed red line}), which effectively quadruples the emission probability relative to the result in a single pulse (coherent enhancement), and this is why we find a single peak in Fig.~\ref{Fig_same_pulse} (b). Similarly, in the range \mbox{$0.14<s<0.16$} ({dotted blue line}) we see that $\cos\Phi_f$ becomes negative, resulting in the destructive interference at the centre of Fig.~\ref{Fig_same_pulse} (d). Finally, if we consider the whole range \mbox{$0.06<s<0.16$} as in Fig.~\ref{Fig_same_pulse} (f), the positive and negative parts of the cosine factor cancel and so interference effects are smoothed out.

\section{Experimental signals}~\label{SECT:EXPT}
\subsection{Photon number}
\begin{figure}[t!]
\includegraphics[width=0.48\textwidth]{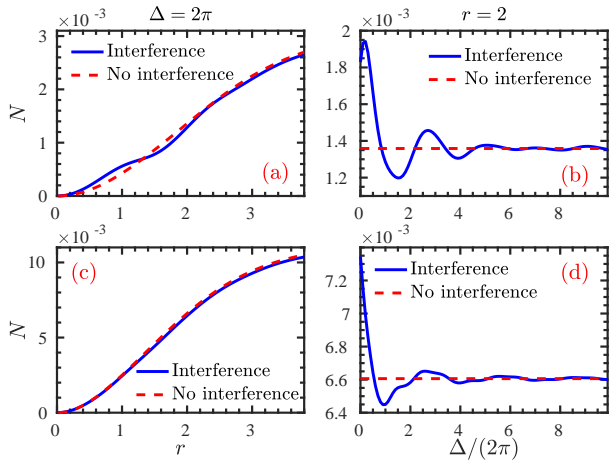}
\caption{Number of photons \Eq{\ref{Eq_NCL_Interference_number_angle_window}}. Left column: fixed pulse separation $\Delta=2\pi$, variable angular window $r$. Right column: $r=2$ fixed, variable separation $\Delta$. In (a) and (b) $\delta s=0.02$, $s_c=0.07$, while in (c) and (d) $\delta s=0.1$, $s_c=0.11$. Other parameters as in Fig.~\ref{Fig_same_pulse}.}
\label{Fig_same_Number}
\end{figure}
Consider measuring the total number of photons impinging on some detector with a finite acceptance angle $\theta_r$ and a given energy resolution $\delta s$.
We calculate the number of photons by integrating the differential probability \Eq{\ref{Eq_NCS_Simplified}} over a range of $s$ and over a square portion, length of side $r$, of the $\ell^\LCperp$ plane centred on the origin, i.e.~
\begin{align}
	N = \int\limits^{s_c+ \delta s/2}_{s_c- \delta s/2}\!\ud s\!\int\limits^{r/2}_{-r/2} \!\ud \ell^x\!\!\!\int\limits^{r/2}_{-r/2} \!\ud \ell^y ~\frac{\ud^3{P}}{\ud s~\ud^2\ell^\LCperp}\,.
\label{Eq_NCL_Interference_number_angle_window}
\end{align}
This works because the length of side $r$ corresponds to the polar angle of the emitted photon via
\begin{align}
\theta_{r} \approx \frac{2\omega_0r}{m\eta}\approx 6.4\times 10^{-2} r~\textrm{mrad}\,,
\end{align}
in which we use the same parameters as in Fig.~\ref{Fig_same_pulse} ($\eta=0.095$, $\omega_0=1.55$eV).  We plot $N$ in Fig.~\ref{Fig_same_Number} for various parameters. The number of received photons increases with acceptance angle, i.e. with~$r$. Fig.~\ref{Fig_same_Number} (a) shows that with a {high} resolution $\delta s =0.02$, this number is modulated by interference, relative to twice the single-pulse result (the incoherent sum), in particular at small acceptance $r<2$. With a {lower} energy resolution, though, interference effects are smoothed out as shown for $\delta s=0.1$ in Fig.~\ref{Fig_same_Number} (c).
 In the right hand column of Fig.~\ref{Fig_same_Number}, we fix the size of the angular window, $r$, and instead vary the separation $\Delta$ between the two pulses. As $\Delta$ increases the photon number oscillates around the result obtained by neglecting interference, and converges to it at {a pulse separation of} around 5 laser cycles. This oscillation in the photon number relies sensitively on the detector energy resolution: for a poor resolution $\delta s=0.1$ as in Fig.~\ref{Fig_same_Number} (d), the oscillations are washed out at smaller $\Delta$.

\subsection{Polarisation effects}
We now consider interference effects which arise when the \textit{polarisation} of the second pulse is different from that of the first. We consider two cases; we either rotate the electric field of the second pulse to be \textit{perpendicular} to that of the first, or to be \text{anti-parallel} to the first. (This second case, which is simply two pulses of opposite sign, {is standard in the consideration of interference effects}~\cite{Akkermans:2011yn}.)

\begin{figure}[t!!!]
 \includegraphics[width=0.49\textwidth]{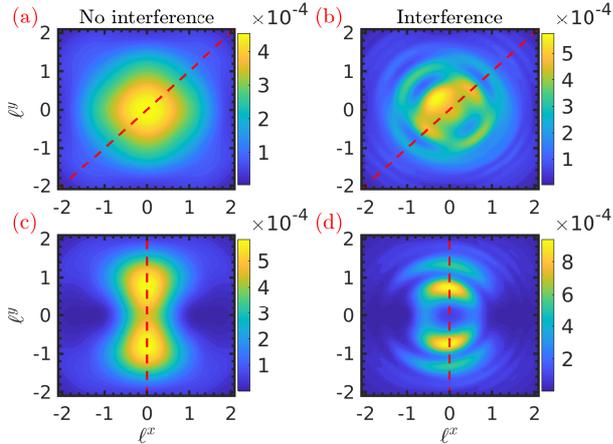}
\caption{Transverse photon distribution \Eq{\ref{WTACTUALF}} of the emitted photon without (left column) and with (right column) interference. Upper panels: the electric fields of the $1$st and $2$nd pulses are polarised in $x$ and $y$ direction respectively. Bottom panels: the electric fields of the pulses are both in $x$ direction, but with opposite sign. The red dashed lines are $\ell^x=\ell^y$ in the upper panels and $\ell^x=0$ in the bottom panels. The other parameters are same as in Fig.~\ref{Fig_same_pulse}.}
\label{Fig_perp_oppo_polarisation}
\end{figure}

Observe that for the example above of two identical pulses polarised in the $x$-direction, the distributions in Fig.~\ref{Fig_same_pulse} were symmetric {about} $\ell^y=0$.
For two pulses with orthogonal polarisation (the first in the $x$-direction, as before, the second in the $y$-direction), we would {na\"{\i}vely} expect symmetry {about} the line $\ell^x=\ell^y$.
However, as shown in Fig.~\ref{Fig_perp_oppo_polarisation} (a) and (b), interference breaks this symmetry.
This asymmetry can be seen in e.g. \Eq{\ref{Eq_NLC_Two_Interference}}, as for perpendicularly polarised pulses we have \mbox{$I_{1}(\ell^x,\ell^y)=I_{2}(\ell^y,\ell^x)$}, and so the combination of interference terms \mbox{$\sim e^{i\Phi_f}I^{*}_{1}I_{2}~+e^{-i\Phi_f} I_{1}I^{*}_{2}$} gives
\begin{align}
\label{asymm}
	\sin\Phi_{f}~(I^{*}_{1}I_{2}-I_{1}I^{*}_{2}) \;,
\end{align}
and similarly for $F_\mu$ and $S$.

Fig.~\ref{Fig_perp_oppo_polarisation} (c) and (d) show similar results for the case where the second pulse is polarised anti-parallel to the first. Without interference, the incoherent result Fig.~\ref{Fig_perp_oppo_polarisation} (c) is exactly the same as in Fig.~\ref{Fig_same_pulse} (a); there is symmetry {about} $\ell^y=0$ because the fields are polarised in $x$ direction, and symmetry {about} $\ell^x=0$ because of the chosen field shape in \Eq{\ref{Eq_field}}. The symmetry {about} $\ell^x$ is lost in Fig.~\ref{Fig_perp_oppo_polarisation} (d) as the interference terms again pick up an asymmetry as in \Eq{\ref{asymm}}.  This can be understood as a consequence of the field shape -- if we exchange $\sin\phi \to \cos\phi$ in (\ref{Eq_field}), the interference fringes would become symmetric {about} $\ell^x=0$. Alternatively, it can be understood as due to the causal aspect of scattering in two pulses~\cite{ilderton2019coherent}; note that if we were to swap the signs of the pulses, then we would obtain, instead of Fig.~\ref{Fig_perp_oppo_polarisation} (d), its reflection in the line $\ell^x=0$. This is equivalent to swapping the \textit{order} of the pulses, since the emitting electron propagates from smaller to larger $\phi$~\cite{brodsky1998quantum,heinzl2001light}.

\subsection{Electron bunch effects}
To observe a robust interference effect, we should consider not a single electron, but a bunch of electrons. We model this by convoluting the differential emission probability with an initial momentum distribution function for the bunch.

Let this distribution be $\rho(\mathbf{p})$, obeying the normalisation condition: \mbox{$\int \ud^3{\bf p}\, \rho(\mathbf{p})=1$} (i.e.~we divide out the total number of electrons in the bunch). We split the total momentum distribution into a piece ${L}(p_z)$ in the laser propagation direction and a piece \mbox{${T}(p_\LCperp)$} in the perpendicular direction, \mbox{$p_{\LCperp}:=(p_x,p_y)$}, so \mbox{$\rho(\mathbf{p}) = {L}(p_z)~{T}(p_\LCperp)$} with
\begin{align}
	{L}(p_{z})=&\frac{1}{\sqrt{2\pi}\sigma_z m}\exp\left[-\frac{(p_{z}-\tilde{p}_z)^2}{2\sigma^2_{z}m^2}\right]\,,\nonumber\\
	{T}(p_\LCperp)=&~\frac{1}{\pi\sigma^2_{\LCperp} m^2}~~\exp\bigg[-\frac{p^2_\LCperp}{\sigma^2_{\LCperp} m^2}\bigg]\,,\nonumber
\end{align}
in which \mbox{$\tilde{p}_z = \langle p_z\rangle$} is the average longitudinal momentum. The momentum spreads in each direction are \mbox{$\langle (p_{z}-\tilde{p}_z)^2\rangle^{1/2}=\sigma_{z}m$} and \mbox{$\langle p^2_\LCperp\rangle^{1/2}=\sigma_{\LCperp} m$}.

{For an electron bunch, it is convenient to take $s$ to refer to the \textit{average} electron momentum $\tilde{p}$, i.e. \mbox{$s = k\cdot l/k\cdot\tilde{p}$}.}
In other words $s$ is scaled by the parameter $\lambda = k\cdot p/k\cdot \tilde{p}$ and now takes values between $0$ and $\lambda$ instead of $0$ and $1$.
The calculations of the functions $I,~F_{\mu},~S$ needed for the emission probability are then as in \Eq{\ref{Eq_phase_integral}}, except that $s$ appearing there should be replaced with $s/\lambda$. We also write $\tilde{\eta}_p=k\cdot \tilde{p}/m^2$.  The differential emission probability then takes the form
\begin{align}
	&\frac{\ud^3{P}}{\ud s~\ud^2 \ell^\LCperp}=\frac{\alpha}{(2\pi\tilde{\eta}_p)^2}\int \!\ud^3 {\bf p}~\rho(\mathbf{p})
\frac{s}{\lambda(\lambda-s)}\nonumber\\
  &~~~~~~~\Big[\big(SI^{*}+S^{*}I-2F\cdot F^{*}\big) g\Big(\frac{s}{\lambda}\Big)-|I|^2\Big]\,.
\label{Eq_NCS_wavepacket}
\end{align}
With this, we can investigate the possibility of experimentally observing interference effects in either the photon energy spectrum or angular spectrum.

A nonzero transverse momentum $p_\LCperp $ of the initial electron enters the differential probability \Eq{\ref{Eq_NCS_wavepacket}} through a linear shift of the normalised photon momentum $\ell^\LCperp$, i.e. \mbox{$\ell^\LCperp\rightarrow \ell^\LCperp \lambda- p^\LCperp/m$}. In order to observe the interference fringes in, say, Fig.~\ref{Fig_same_pulse} (b), the shifts introduced by $p^\LCperp$ should be smaller than the gaps between the interference fringes. From Fig.~\ref{Fig_same_pulse}, we can see that the gap between fringes is much smaller than the transverse spread induced by the field, \mbox{$\ell^x\approx \ell^y \approx \xi$}.  The average shift due to the bunch, on the other hand, can be estimated as being equal to the transverse momentum spread $\sigma_t$. This yields the qualitative condition that the transverse spread of the beam should obey $\sigma_t\ll \xi$. Neglecting the longitudinal distribution, so \mbox{$L(p_{z})=\delta(p_{z}-\tilde{p}_z)$}, and using the same parameters as in Fig.~\ref{Fig_same_pulse} (b) (\mbox{$\tilde{p}_z=8~\textrm{GeV}$} and \mbox{$\tilde{\eta}_p=0.095$}), we find that to reproduce the interference fringes in Fig.~\ref{Fig_same_pulse} (b), we would have to reduce the transverse momentum spread in the beam to \mbox{$\sigma_{\LCperp}=\xi/20$}, corresponding to an angular divergence in the bunch of \mbox{$\Theta = 2\sigma_{\LCperp} m/\tilde{p}_z\approx 1.3\times10^{-2}~\textrm{mrad}$}, which is orders of magnitude smaller than recent experimental results~\cite{leemans14,PRL2019Petawatt}. Therefore, the transverse spread of an electron bunch would seem to make it challenging to observe interference in the angular distribution of the emitted photons.

\begin{figure}[t!!]
 \includegraphics[width=0.45\textwidth]{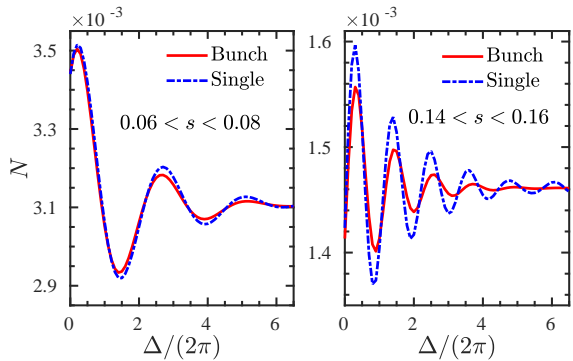}
\caption{Oscillation of the photon number $N$ with the change of the pulse separation. Dash-dotted lines: single-electron results; Solid lines: results from a electron bunch with normalized momentum distribution, see the discussions about the momentum distribution in the text. The other parameters are same as in Fig.~\ref{Fig_same_pulse}.}
\label{Fig_wavepacket_number}
\end{figure}

However, we will now see that it may be possible to detect interference in the photon energy spectrum, which we obtain from \Eq{\ref{Eq_NCS_wavepacket}}, by integrating out the transverse degrees of freedom,
\begin{align}
\label{Eq_NCS_wavepacket_long}
&\frac{\ud{P}}{\ud s}=\frac{\alpha}{(2\pi\tilde{\eta}_p)^2}\int\limits^{\infty}_{s} \frac{\ud\lambda}{\lambda}\int \ud^2p_\LCperp \frac{p^{0} \rho({\bf p})  s}{\lambda(\lambda-s)}\\
&\int \ud^2 \ell^{\LCperp} \Big[\big(SI^{*}+S^{*}I-2F\cdot F^{*}\big) g\Big(\frac{s}{\lambda}\Big)-|I|^2\Big]\;. \nonumber
\end{align}
Note that due to the form of $I,~F_{\mu},~S$, the only dependence on the transverse electron momentum $p_\LCperp$ in \Eq{\ref{Eq_NCS_wavepacket_long}} lies in the factor $p^{0} \rho(\mathbf{p})$.
We are interested in small-angle collisions, hence the dependence of $p^{0}$ on $p^{\LCperp}$ is negligible and the bunch transverse distribution can be integrated out: $\int \ud^2p_\LCperp {T}(p_\LCperp)=1 $. Hence the photon energy spectrum is determined by the longitudinal momentum distribution of the bunch.

Multi-GeV electron beams with very narrow energy spread and angular divergence are available to experiment~\cite{leemans14,PRL2019Petawatt}. We take the average momentum in the beam to be \mbox{$\tilde{p}_\mu=(\tilde{p}_{0},0,0,-\sqrt{\tilde{p}_{0}^2-m^2})$} with the energy \mbox{$\tilde{p}^{0}=8~\textrm{GeV}$}, and widths \mbox{$\sigma_z=3\%~\tilde{p}_{0}/m$} and \mbox{$\sigma_t=10^{-4}~\tilde{p}_{0}/m$} corresponding to an energy spread of $6\%$ and an angular divergence of $\Theta=0.2$ mrad following Ref.~\cite{PRL2019Petawatt}. In Fig.~\ref{Fig_wavepacket_number}, we show the number $N$ of photons produced, in a given energy range, as a function of the pulse separation $\Delta$. Similar to the single-particle result in which the electron possesses a definite momentum $p=\tilde{p}$, there is a pronounced oscillation of the photon number even taking into account bunch effects. This evanesces with the increase of the pulse separation. With a broad longitudinal spread in the bunch, the amplitude of the interference oscillations reduces and converges to the incoherently summed result at a shorter pulse separation than the single-electron result.

\section{Discussion and Conclusion}~\label{SECT:CONCS}

We have investigated interference effects in the collision of an energetic electron bunch with a sequence of two laser pulses. Interference {effects} are present in both the scattered electron and emitted photon spectra, and we have focussed on the latter. The potential observation of interference effects depends both on the properties of the electron bunch and on the detector resolution in the experiment.

Accounting for both, we have seen that interference effects may be observed experimentally in the oscillation of the {detected number of photons} as a function of varying the pulse separation. This would require multi-GeV electron bunches with narrow energy spread, which are available from laser plasma acceleration~\cite{PRL2019Petawatt}. For higher energy resolution, interference effects persist for longer and stronger laser pulses, as well as for larger pulse separations. In future work one could incorporate additional interference effects into the analysis by considering e.g.~coherent emission from bunches of particles~\cite{BERRYMAN1996526,PhysRevLett.121.010402,PhysRevD.99.035048}.
   
We finally remark that if the transverse divergence of the electron bunch could be reduced to be around one order of magnitude smaller than the dimensionless laser intensity, so $\sigma_t\ll\xi$, then it would be possible to also observe interference fringes in the transverse photon distribution. From \Eq{\ref{Eq_NLC_Interference_same_prob}}, we know that the interference fringes appear at the positions where $\Phi_f=2n\pi$, $n$ is an arbitrary integer. If we could measure the position of the interference fringes, we could then infer the intensity of the laser pulse, for a given field shape, from \Eq{\ref{Eq_interference_phase}}.

\subsection*{Acknowledgments}
 The authors are supported by the EPSRC, Grant No.~EP/S010319/1.

\section*{References}

\bibliographystyle{apsrev}
\providecommand{\noopsort}[1]{}

\end{document}